# Anomalies of undercoordinated water molecules

Chang Q Sun

Ecqsun@ntu.edu.sg

Nanyang Technological University, Singapore

**Molecular undercoordination shortens and stiffens the H-O bond but lengthens and softens the O:H nonbond simultaneously associated with O 1s energy entrapment and nonbonding electron dual polarization, which dictates behavior of water and ice dominated by undercoordinated molecules, such as droplets, bubbles, defects, skins, etc.**

1 Anomalies

Undercoordinated water molecules are referred to those with fewer than four nearest neighbors as they occur in the bulk interior of water [1-6]. Molecular undercoordination occurs in the terminated O:H-O bonded networks, in the skin of a large volume of water and ice, in hydration shells, molecular clusters, ultrathin films, snowflakes, clouds, fogs, nanodroplets, nanobubbles, and in the gaseous state. With the involvement of molecular undercoordination (CN < 4) as a new degree of freedom, water molecules perform even more strangely, deriving with the following mysteries:

1)  Low density yet mechanically stiffer and thermally more stable.
2)  Ice is most slippery of ever known – covered by liquid water?
3)  Water skin is elastic, hydrophobic, and tough – covered by solid?
4)  Undercoordination not only elevates the melting point but also depresses the freezing point.

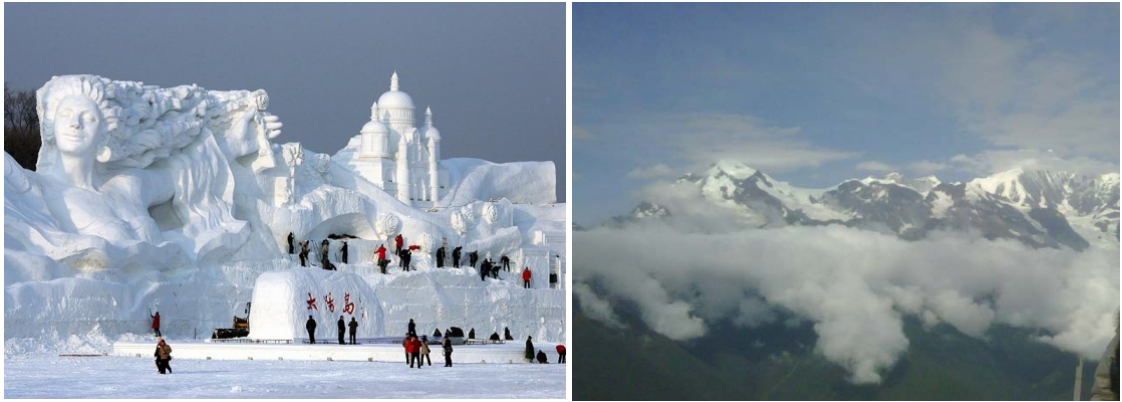

Figure 1. Skins consisting undercoordinated molecules (with fewer than four nearest neighbors as they are in the bulk) brought excessive anomalies to water and ice. (a) A snow sculpture at the Harbin International Ice and Snow Sculpture Festival held on December, 2007 – 'Romantic Feelings' in Harbin, China. (Credit: Timeea Vinerean 2011, Public domain.) (b) Could, fog, and snow photo taken by Yi Sun at Meili Mountains, Yunnan, China, 2010.

2 Reasons

These anomalies arise simply from molecular undercoordination or molecules with fewer than four nearest neighbors as they are in the bulk [7, 8], see Figure 2:

1) Molecular undercoordination shortens the H-O bond and lengthens the O:H nonbond with dual polarization that raises the hydrophobicity, viscoelasticity, and repulsivity.
2) H-O bond stiffening raises its phonon frequency $\omega_H$, Debye temperature $\Theta_{DH}$, O 1s binding energy $E_{1s}$, and energy $E_H$ for H-O atomic dissociation.
3) O:H nonbond elongation lowers its phonon frequency $\omega_L$, Debye temperature $\Theta_{DL}$, and the $E_L$ for molecular dissociation.
4) The $\omega_H$ elevation and the $\omega_L$ depression offset their respective specific heat curve and disperse the quasisolid phase boundaries, elevating the $T_m$ for melting and depressing the $T_N$ for ice nucleation.

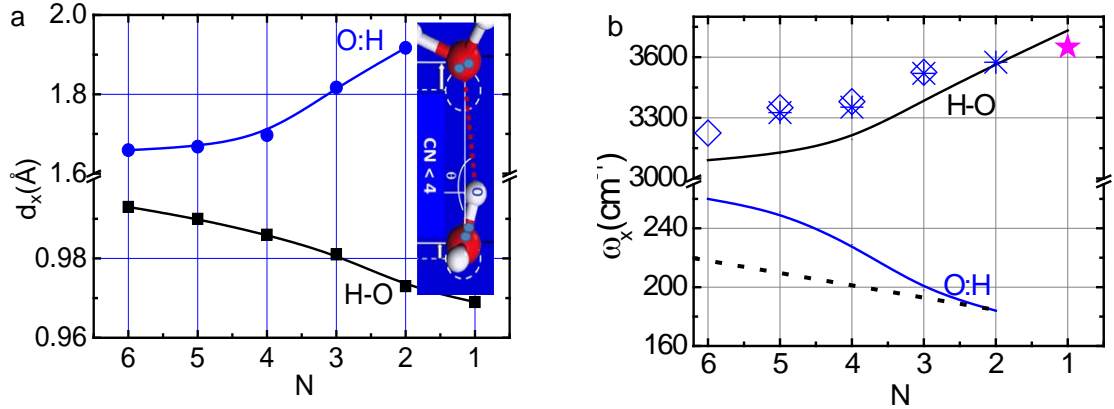

Figure 2. Cluster size (N) dependence of the segmental (a) length $d_x$ and (b) phonon frequency $\omega_x$ of the O:H-O bond in $(H_2O)_N$ clusters [7]. Broken line suggest correction with respect to measured bulk O:H vibration frequency 220 cm$^{-1}$. Scattered symbols are experimental results [9-12].

## 3    History background

??

## 4    Quantitative evidence

### 4.1    BOLS theory

Bonds in skins of metals, alloys, semiconductors, insulators and nanostructures contract globally with respect to the underlying bulk material. The spacing between the first and the second atomic layers of these systems contracts by 10–14% relative to the bulk geometry. For nanostructures, the extent of contraction is greater; the relaxation extends radially into deeper layers [13, 14]. For the one-dimensional atomic chains and edges of the two-dimensional graphene ribbons, bonds contracts by up to 30%. Sun undercoordination induced bond contraction takes the full responsibility for the unusual behavior of the undercoordinated adatoms, point defects, and nanostructures of different shapes, including the size-dependence of the known bulk properties and the size-induced emergence of properties that the bulk parent never shows [15]. Atomic undercoordination induced local bond contraction and the associated energy change follow the bond order-length-strength (BOLS) correlation premise [15, 16].

The curvature $K^{-1}$ (K being the dimensionless form of size is the number of atoms lined along the radius of a spherical dot) dependence of the effective atomic coordination ($z_i$), bond length ($d_i$), charge density ($n_i$), energy density (proportional to the elastic modulus, $B_i$), and the potential trap depth ($E_i$) in the ith atomic site follow the relationships [15]:

$$\begin{cases} z_1 = 4(1-0.75K^{-1}); z_2 = z_1 + 2; z_3 = 12 & \text{(Effective CN)} \\ d_i = C_i d_b = 2d_b\left[1+\exp\left((12-z_i)/(8z_i)\right)\right]^{-1} & \text{(Bond contraction coefficient)} \\ E_i = E_b C_i^{-m} & \text{(Bond energy; Potential trap depth)} \\ B_i = B_b C_i^{-(m+3)} & \text{(Energy density; Elastic modulus)} \\ n_i = n_b C_i^{-m} & \text{(Charge density)} \end{cases}$$

(1)

The bond nature index $m$ varies approximately from one to four when turn a bond from metallic to covalent. The subscript $i$ denotes the $i^{th}$ atomic layer counted from the outermost inward and the subscript $b$ denotes the corresponding bulk values. Therefore, one can focus on the energetic and electronic behavior of the skin bonds of a certain weightage over the entire object while keeping in mind that the core interior ($i > 3$) of a nanostructure remains its bulk attribute [13, 17]. Figure 3 formulates the atomic CN–resolved bond-length contraction coefficient in comparison with the measured data for atomic chains, liquid and solid skins, Au nanoparticles, graphite and carbon nanotubes, as discussed above and documented in [15].

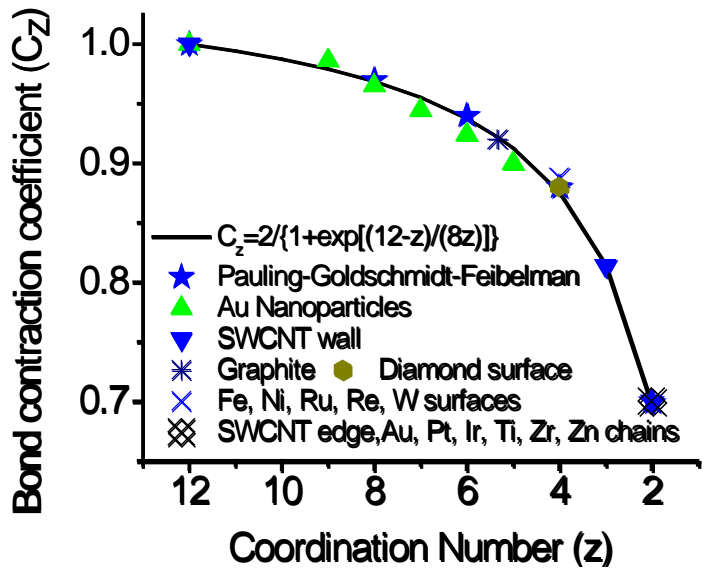

Figure 3. Atomic CN-resolved bond contraction coefficient in comparison with measured data as documented in [15].

Water molecules with fewer than the ideal four nearest neighbours in the bulk should follow the BOLS notion. However, the involvement of lone-pair ':' interaction and O-O repulsion prevents the O:H and the

H-O from following the BOLS notion simultaneously because the lone pairs screen an H$_2$O molecule from intercation of its neighbours. The O:H-O binding energy disparity means that the stronger H-O bond serves as the 'master' to contract by a different amount from what the BOLS notion predicts, as Figure 2a inset shows. The contraction of the H-O bond lengthens and softens the 'slave' O:H nonbond by Coulomb repulsion, with a dual process of polarization. The phonon frequency $\omega_x$ ($x$ = L for the O:H nonbond; $x$ = H for the H-O bond) characterizes the stiffness of the respective segment in the following [16],

$$\omega_x^2 \propto \left.\frac{\partial^2 u_x(r)}{\mu_x \partial r^2}\right|_{r=d_x} \propto \frac{E_x}{\mu_x d_x^2} \propto \frac{Y_x d_x}{\mu_x}$$

The stiffness is the production of the elastic modulus $Y_x$ and the segmental length $d_x$ and the $Y_x$ is poroportional to the local energy density $E_x d_x^{-3}$. $\mu_x$ is the reduced mass of the vibrating dimer.

It is universally true that one segment of the O:H-O bond will be stiffer if it becomes shorter; it will be softer if it becomes longer [18]. Therefore, the phonon frequency shift $\Delta\omega_x$ tells directly the variation in length, strength and stiffness of the particular segment subjected to an applied stimulus. Because of the Coulomb repulsion, $\omega_L$ and $\omega_H$ shift such that if one becomes stiffer, the other will be softer.

However, the XRD radial distribution function gives only the statistic mean of the O-O distance without discriminating H-O contraction from O:H elongation for the O:H-O bond between undercoordinated molecules. Electron spectroscopy (UPS, XPS, XAS, XES) could not even discriminates the O:H nonbond contribution from the H-O contribution to the electronic binding energy shift. Being only 3% of the H-O bond energy (~4.0 eV), the O:H energy (~0.1 eV) contributes negligibly to the Hamiltonian that determines the electron binding energy shift. Therefore, phonon spectrometrics is most powerful to investigating systems like water and ice to monitor phonon frequency shifts of multiple short-range interactions.

4.2    Structure geometric order

Figure 4 and Table 1 feature the bond geometry, ∠O:H-O containing angle, O:H-O bond segmental length and energy for the optimal (H$_2$O)$_N$ structures [7]. The (H$_2$O)$_N$ structure varies from chain (N = 2), ring (N = 3-5), cage (N = 6-10) to solid clusters (N = 12 – 20). N = 6 derives the book, prism, and cage like structures of the same binding energy [19]. The O:H-O configuration holds for any clustering geometry despite the ∠O:H-O containing angle varying from 160 to 177 ° and segmental lengths. The effective CN of the H₂O also varies from situation to situation. For the same N value, the CN varies with the dimensionality of chains, rings, cages, and solid clusters. Therefore, the O:H-O configuration and the associated containing angle and the segmental lengths are the key identities in all possible geometrical configurations of water and ice.

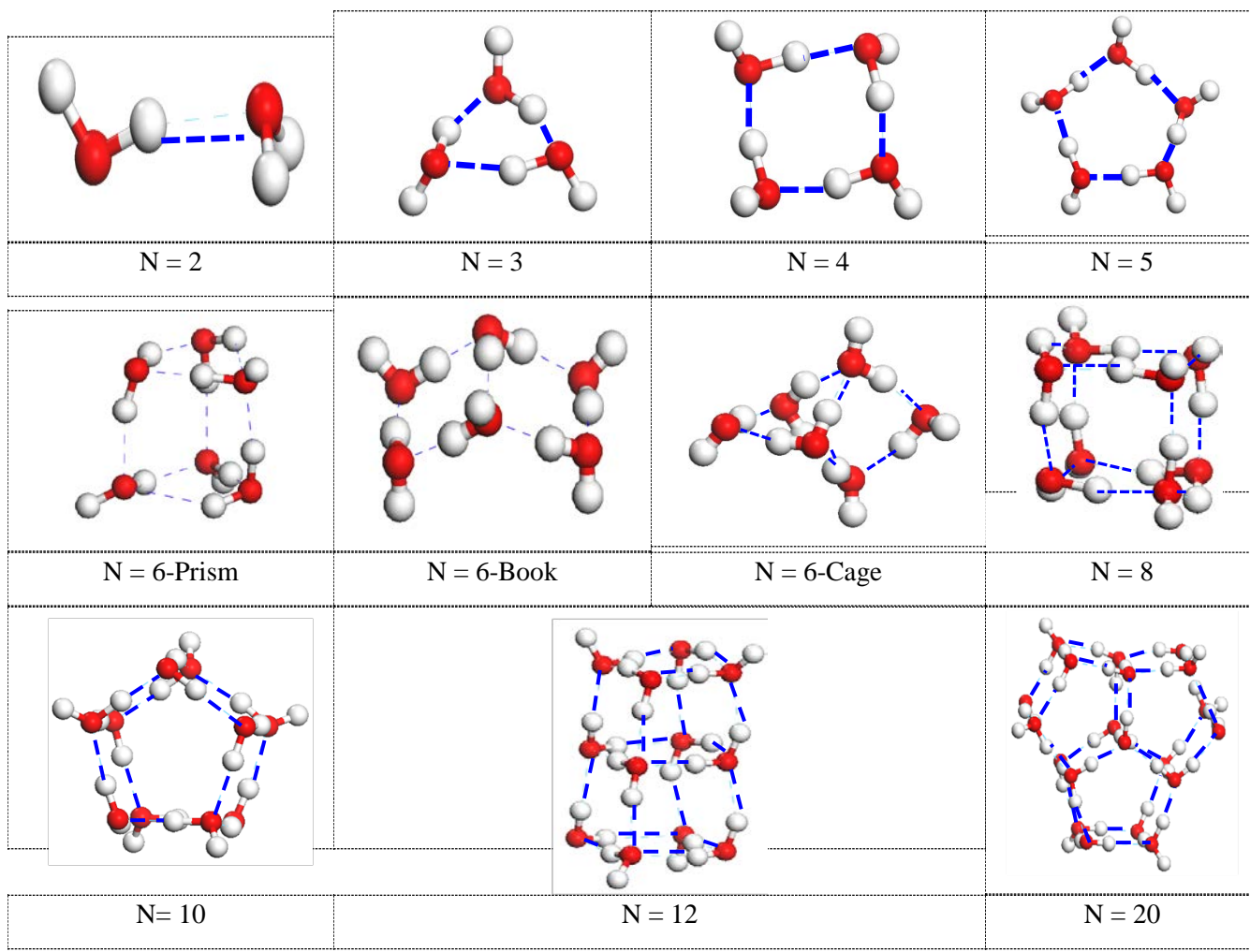

Figure 4. DFT optimized (H$_2$O)$_N$ crystal structures showing the chain like (N = 2), ring-like (N < 6), cage-like (N = 6-10), and solid clusters (N = 12-20). N = 6 creates three structures of nearly identical binding energy [7, 19].

Table 1. DFT-derived segmental length $d_x$, total electronic binding energy $E_{Bind}$ (-eV), and segmental $E_x$, of (H$_2$O)$_N$ clusters*. References contain experimental results [7].

|  | N | $d_H$(Å) | $d_L$(Å) | θ(°) | $d_{OO}$(Å) | $E_{Bind}$(eV) | $E_H$(eV) | $E_L$(eV) |
|---|---|---|---|---|---|---|---|---|
| Monomer | 1 | 0.969 | - | - | - | 10.4504 | 5.2252 | - |
| Dimer | 2 | 0.973 | 1.917 | 163.6 | 2.864 | 21.0654 | 5.2250 | 0.1652 |
| Trimer | 3 | 0.981 | 1.817 | 153.4 | 2.730 | 31.8514 | 5.2238 | 0.1696 |
| Tetramer | 4 | 0.986 | 1.697 | 169.3 | 2.672 | 42.4766 | 5.2223 | 0.1745 |
| Pentamer | 5 | 0.987 | 1.668 | 177.3 | 2.654 | 53.1431 | 5.2208 | 0.1870 |
| book | 6 | 0.993 | 1.659 | 168.6 | 2.640 | 63.8453 | 5.2194 | 0.2020 |
| Cages | 6 | 0.988 | 1.797 | 160.4 | 2.748 | – | – | – |
|  | 8 | 0.992 | 1.780 | 163.6 | 2.746 | – | – | – |
|  | 10 | 0.993 | 1.748 | 167.0 | 2.725 | – | – | – |
| Clusters | 12 | 0.992 | 1.799 | 161.7 | 2.758 | – | – | – |
|  | 20 | 0.994 | 1.762 | 165.4 | 2.735 | – | – | – |
| Bulk | Ih | 1.010 | 1.742 |  |  |  | 3.97[20] |  |
| Bulk [21, 22] | 4°C | 1.0004 | 1.6946 | 160.0 | 2.6950 | 536.6[23] |  | 0.095[24] |

*The total energy of a cluster is that required to excite all the electrons to the vacuum level; the binding energy $E_{Bind}$ is the energy required to combine atoms together to form a cluster:

$E_{Bind} = E_{cluster} - \sum E_{atom} = \sum E_{bond}$. For N = 1, there are two H-O bonds only. Thus, the H-O bond energy is one-half of the $E_{Bind}$: $E_H(1) = E_{Bind}(1)/2$. For N = 2 to 6: $E_H(N) = E_H(1) - \alpha(d_H(N) - d_H(1))^2$, where α is the elastic constant.

The ∠O:H-O containing angle or the ∠O-H-O angle and the segmental length and energy determine the geometries of snowflakes and ice crystals, see Figure 5. Any pertrubation by bioelectrionic stimulus like emiotion [25], mechanical exitation like sound tones and frequencies, and thermal pulse or fluctualtional signinals will mediate the growth modes -shape and geometry of ice flakes as the O:H nonbond with binding energy in the order of 10s meV is too sensitive to the said stimulus. Room temperature at 300 K corresponds to 25 meV only.

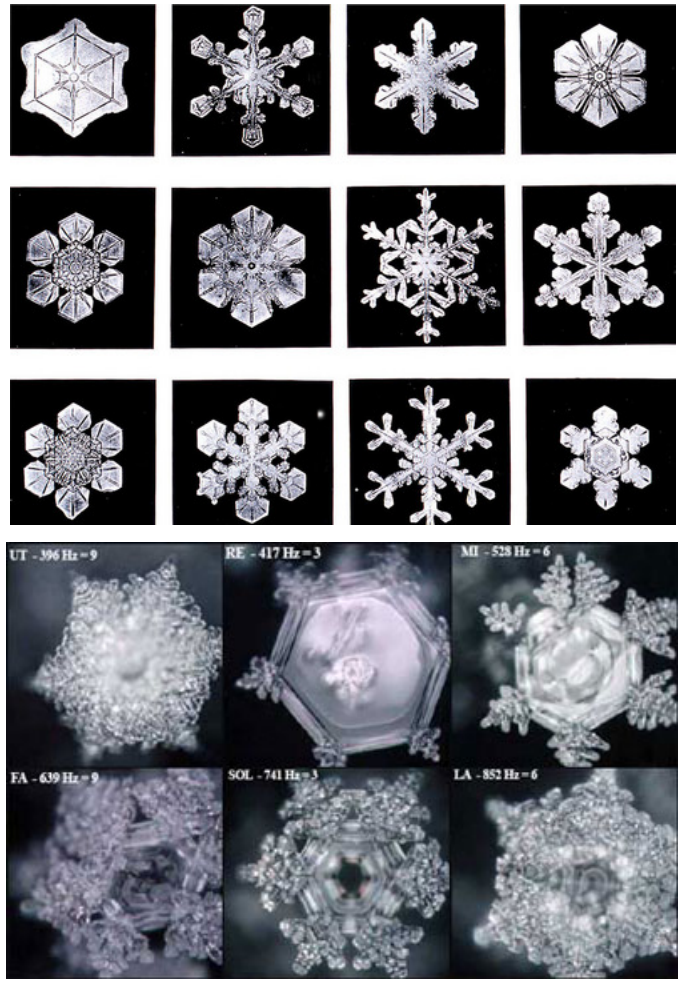

Figure 5. Geometries of snowflakes (upper) that have fascinated many eminent scientists and philosophers such as René Descartes, Johannes Kepler, and Robert Hooke, but the man who literally devoted his entire life to showing us the diversity and beauty of snowflakes is American Wilson A. Bentley (February 9, 1865 – December 23, 1931). (Credit: Timeea Vinerean 2011, Public domain.) Ice geometries grow under music (Credit: Masaru Emoto [25]).

Masaru Emoto, a doctor of alternative medicine from Japan, has conducted many experiments on ice crystals and found that words and music have both positive and negative effects. Emoto has published multiple books on his observations. One of them is *The Hidden Messages in Water*. Emoto firstly put water into containers and then labelled them with written or typed thoughts and feelings. He took immediately microscopy photos of the frozen water crystals during freezing. Beautiful words such as love and gratitude derived beautiful crystals, while misshapen and distorted crystals were formed when slang or swear words were used. Polluted water formed ugly crystals, but after polluted water was prayed over, prettier water crystals resulted. Figure 5 showed some examples of the ice crystals.

Such amazing findings show that words, both written and spoken, can actually have an effect on water. Emoto also did these water crystal experiments with music. He put water between two speakers and turned on a specific piece of music for several hours. Then the water was frozen and photographed. Classical music such as those by Beethoven and Bach resulted into beautiful water crystals. Tibet Sutra and the Kawachi Folk Dance all produced pretty water crystals. However, rock music did not have the same results - instead, rings of cracks showed up when the water exposed to rock music was frozen.

Since 70% of our bodies are made up of water, we can apply the healing properties of classical music to our own bodies. Every piece of music has different frequencies, and these frequencies can reach different parts of our body to assist our immune systems and minds to achieve healing and other positive effects. Emoto said that he sees energy as vibrations moving through matter. These vibrations, called hado by Emoto, translate as wave motions or vibrations. For example, listening to
The Moldau by Bedrich Smetana can reduce irritability and energize the lymphatic tissues of the body. Emoto attributes this to the hado in the music that can cancel out the hado of irritability. The music also resonates with the hado of lymphatic system. Other examples include listening to The Blue Danube by Johann Strauss II will revitalize your central nervous system, while listening to the opera and ldquo; Lohengrin and rdquo; by Richard Wagner will remove the hado of self-pity from your thoughts. Emoto's observations showed indeed that the O:H nonbond is very sensitive to the external stimulus even at extremely low levels.

4.3     O:H-O segmental length, energy, and mass density

X-Ray absorption spectrosocpy revealed that the skin O-O distance for water expands by 5.9% or more by inter-oxygen Coulomb repulsion, compared to a 4.6% contraction of the skin O-O distance for liquid methanol [26], which differentiates the surface tension (should be compression instead) of 72 mN/m for water from 22 mN/m for methanol. The O-O distance in the skin and between a dimer is about 3.00 Å; the O-O distance in the bulk varies from 2.70 [27] to 2.85 Å [28], depending on experimental conditions. The ideal O-O distance at 4 °C is 2.6950 Å [22]. The volume of water confined in 5.1 and 2.8 nm sized $TiO_2$ pores expands by 4.0% and 7.5%, respectively, with respect to the bulk water [29]. MD calculations also reveal that the $d_H$ contracts from 0.9732 Å at the center to 0.9659 Å at the skin of a free-standing water droplet containing 1000 molecules [30]. X-ray scattering, neutron reflection, and sum-frequency vibrational spectroscopy suggested that the boundary water layer in the vicinity of hydrophobic surface consists of a ~0.5 nm depletion layer with a density of 0.4 g/cm$^3$ and a considerable amount (25-30%) of water molecules with free OH groups [31]. The 0.4 g·cm$^{-3}$ density correpsonds to O-O distance of $d_{O-O}$ = 3.66 Å.

Figure 6a and Table 1 show O:H-O bond segmental relaxation as a function of N for the $(H_2O)_N$ clusters derived from calculations using the PW and OBS algorithms [7]. As N is reduced from 24 (an approximation to the bulk) to two (a dimer), the H-O bond contracts by 4% from 0.101 nm to 0.097 nm, and the O:H bond expands by 17% from 0.158 to 0.185 nm, according to the OBS derivatives. Figure 6b plots the $N$-dependence of the O-O distance that expands by 8%, when the $N$ is reduced from 20 to 3, which is compatible to the value of 5.9% measured in the water skin at 25 °C [26]. This cooperative relaxation expands the O-O distance by 13% and lowers the density by 30% for the dimer. The monotonic relaxation profiles for the $d_x$ at N ≤ 6 will be the focus in the subsequent discussions without rendering the generality of conclusion.

Consistency of the BOLS predictions with experimental [3, 4] and numerical [7] observations confirms the following:

1) The H-O bond shortening (lengthening) is always coupled with the O:H lengthening (shortening), irrrespective of the algorithm used, which is evdence of the expected O:H-O bond cooperativity – one segment contracts and the other must expand because Coulomb coupling.
2) The non-monotonic change of $d_x$ results from the effective CN that varies not only with the number of molecules N but also with the geometrical configuration of the $(H_2O)_N$ cluster. The effective CN of a ring-like cluster is smalller than that of a cage for the same N value.
3) Molecular undercoordination increases the $E_H$ and reduces the $E_L$ in magnitude, as the BOLS notion predicts.
4) The inconsistent results due to different algorithms suggest that one should focus more on the trend and the natural origin than on the accuracy of the derived values. Numerical derivatives serve as useful references for concept verification.

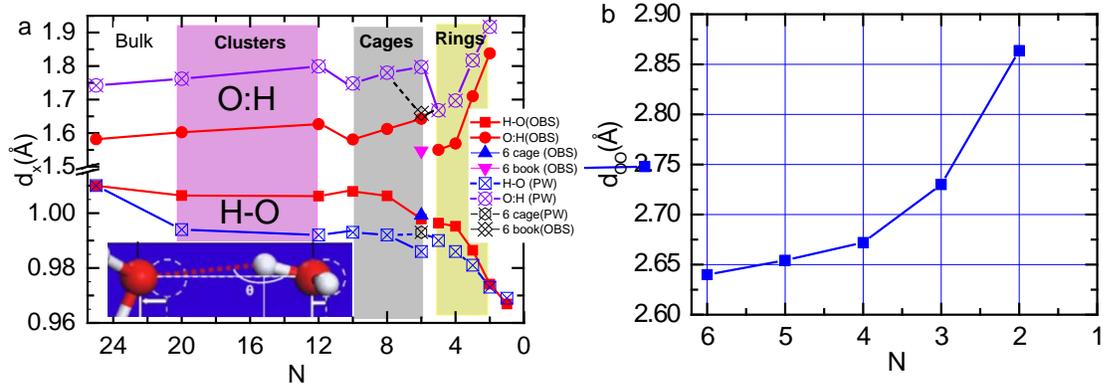

Figure 6. Cluster size-dependence of (a) $d_x$ in the $(H_2O)_N$ clusters optimized using the PW [32] and the OBS [33] methods. N = 6 gives the 'cage', 'book', and 'prism' hexamer structures, with almost identical binding energies. The non-monotonic tread stems from the effective molecular CN that also changes with geometrical configuration. N dependence of (b) the $d_{OO}$ for N = 2-6 gives the mass density in the form of $\rho \propto (d_{OO})^{-3}$. (Reprinted with permission from [7].)

4.4    H-O bonding electron entrapment and densification

Following the same trend as 'normal' materials, molecular undercoordination imparts to water local charge densification [34-39], binding energy entrapment [35, 40-42], and nonbonding electron polarization [37]. Figure 7a shows that the O 1s level shifts more deeply from the bulk value of 536.6 eV to 538.1 eV and 539.7 eV when move from bulk water to its skin to monomer in gaseous phase [23, 43, 44]. The O1s binding energy shift is a direct measure of the H-O bond energy and the contribution from the O:H nonbond is negligibly small.

Atomic undercoordination lowers the atomic cohesive energy, a product of the bondenergy and th eatomic CN(z), $zE_z$, that determines the thermal stability of 'normal' materials in general. The energy necessary for dissociating a $(H_2O)_N$ cluster into $(H_2O)_{N-1} + H_2O$ increases, conversely, when the cluster size is reduced to a trimer (Figure 7b) [45], which conflicts with the traditional understanding of 'normal' material behavior.

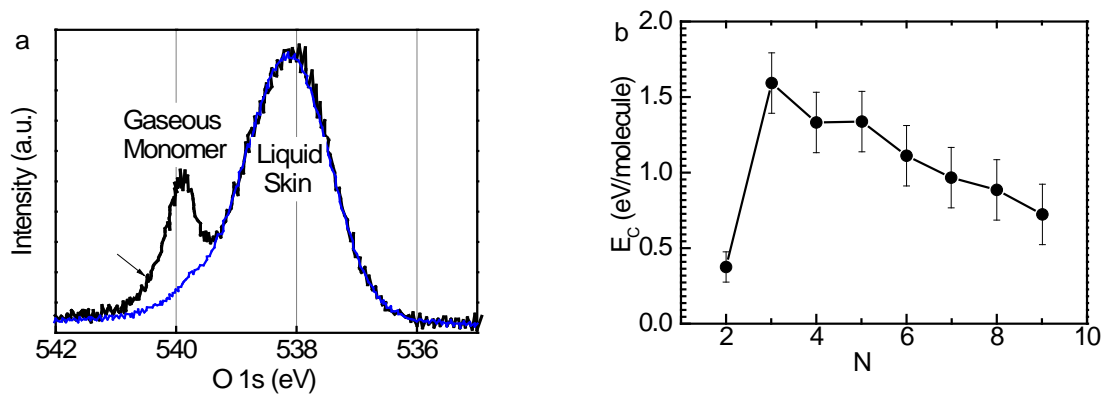

Figure 7. (a) XPS O1s spectra of water containing emission from the liquid skin at 538.1 eV and from the gaseous phase at 539.9 eV (reprinted with permission from [44]); (b) energy required for dissociating a $(H_2O)_N$ cluster into $(H_2O)_{N-1} + H_2O$ (1 kJ/mol = 0.02 eV/molecule). (Reprinted with permission from [45].)

4.4     Nonbonding electron dual polarization

An ultrafast liquid-jet UPS [37], shown in Figure 8, resolved the vertical bound energies (being equivalent to work function) of 1.6 eV and 3.3 eV for the solvated electrons in the skin and in the bulk interior of water solution, respectively. The bound energy decreases with the number $n$ of the $(H_2O)_n$ clusters toward zero, which evidences that molecular undercoordination substantially enhances nonbonding electron polarization, as illustrated in Figure 8d [46].

The nonbonding electrons are subject to dual polarization when the molecular CN is reduced [6]. Firstly, H-O bond contraction deepens the H-O potential well and entraps and densifies electrons in the H-O bond and those in the core orbitals of oxygen. This locally and densely entrapped electrons polarize the lone pair of oxygen from the net charge of -0.616 e to -0.652 eV accoring to DFT calculation for ice skin [8]. The increased charge of O ions further enhances the O-O repulsion as the second round of polarization. This dual polarization rasies the valence band energy up, as shown in Figure 8d. Further reduction of cluster size, or the molecular CN, enhances this dual polarization, resulting observations in Figure 8c – cluster trend of the solvate electron polarization. Therefore, electron dipoles formed on the flat and the curved skins enhaces such polarization, which creates the repulsive force, making liquid water hydrophobicity and ice slippery.

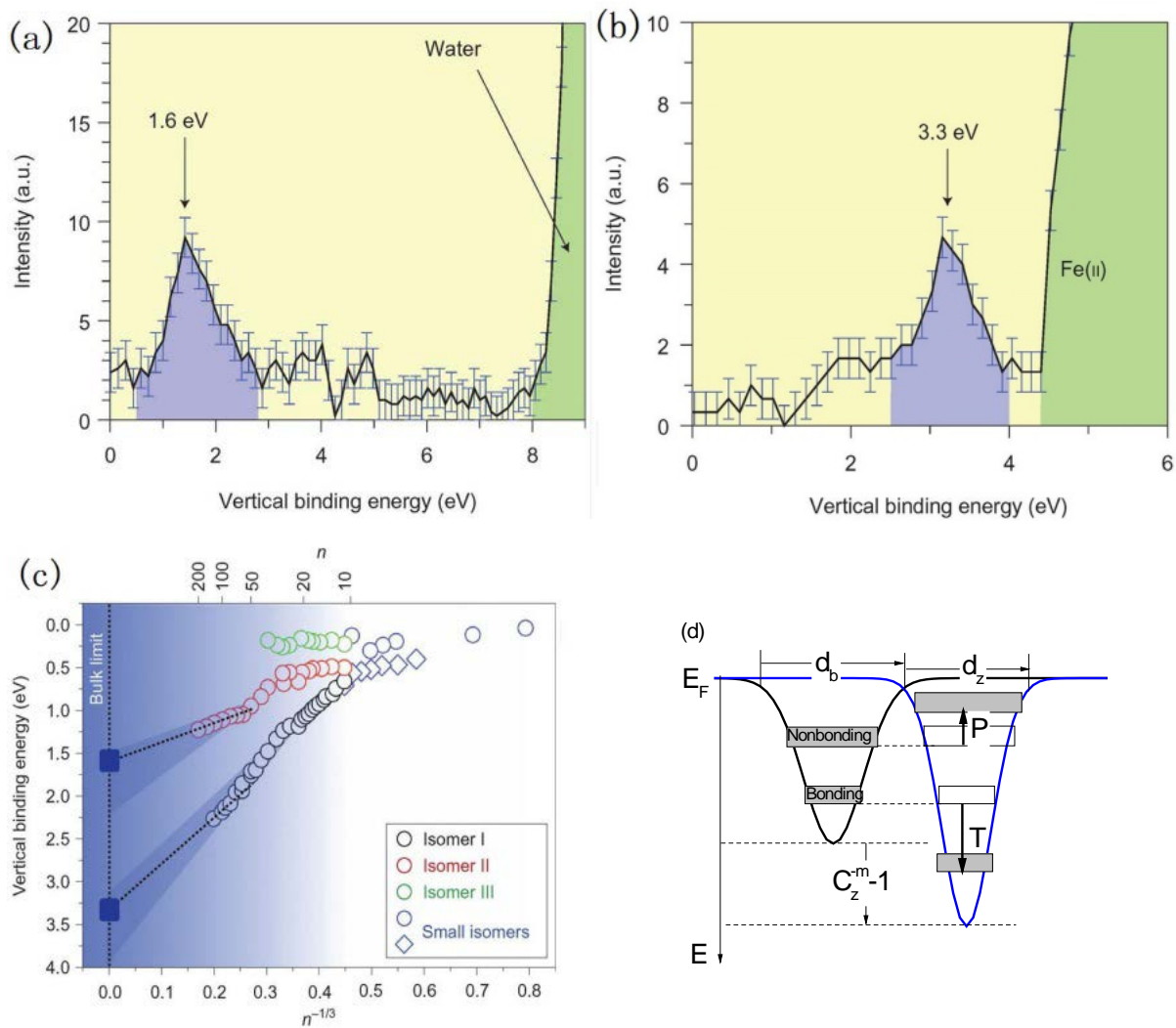

Figure 8. Molecular undercoordination polarizes nonbonding electrons. The vertical bound energy for solvated electrons drops to (a) 1.6 eV in the skin from (b) 3.3 eV in the bulk of the liquid water. (c) The bound energy of solvated electrons in the skin and in the bulk reduces further with the number $n$ of $(H_2O)_n$ clusters toward zero. (d) Nonbonding electron polarization (NEP) theory indicates that molecular undercoordination polarizes nonbonding electrons in two rounds by the densely entrapped H-O bond and O-O repulsion [6].(Reprinted with permission from [37].)

4.5    Cooperative phonon relaxation

Normally, the loss of neighboring atoms softens the phonons of 'normal' materials such as diamond and silicon except for the G mode (1550 cm$^{-1}$) for graphene [47] and the $A_1$ mode (141 cm$^{-1}$) for TiO$_2$ [48] as these two modes arise from dimer vibration independent of atomic CN. However, water molecular undercoordination stiffens its stiffer $\omega_H$ significantly [49, 50]. The $\omega_H$ has a peak centered at 3200 cm$^{-1}$ for bulk water, and at 3450 cm$^{-1}$ for the skins of water ice [51]. The $\omega_H$ for gaseous molecules is around 3650 cm$^{-1}$ [9, 11, 52, 53]. The $\omega_H$ shifts from 3200 to 3650 cm$^{-1}$ when the N of the $(H_2O)_N$ cluster drops from 6 to 1 (Figure 9a) [10-12]. Encapsulation by Kr and Ar matrices lowers the $\omega_H$ slightly by 5–10 cm$^{-1}$ due to the involvement of interface interaction [12]. Size-induced $\omega_H$ stiffening also occurs in large molecular clusters [53] (see Figure 9b). When N drops from 475 to 85, $\omega_H$ transits from the dominance of the 3200 cm$^{-1}$ component (bulk attribute) to the dominance of the 3450 cm$^{-1}$ component (skin attribute) [54]. The high frequency at approximately 3700 cm$^{-1}$ corresponds to the vibration of the dangling H-O radicals, with possible charge transportation in the skin of water and ice [55, 56].

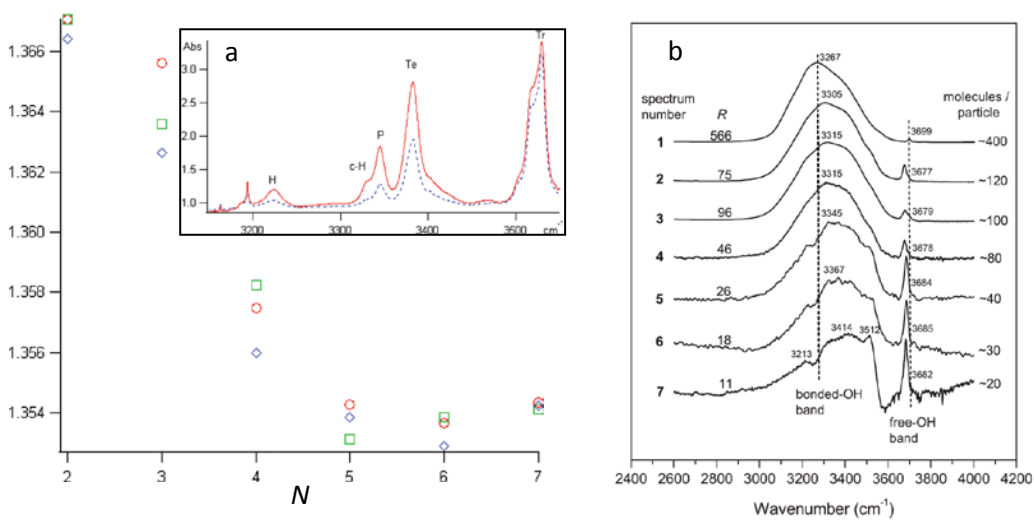

Figure 9. Size-dependent $\omega_H$ of (a) $(H_2O)_N$ clusters (in the frequency ratio of H-O/D-O) and (b) large clusters. Line (2) corresponds to a dimer [57], (3, Tr) to a trimer [58], (4, Te) to a tetramer, (5, P) to a pentamer, (6, c-H) corresponds to a cyclic hexamer, (7, H) corresponds to a cage hexamer. Red circles correspond to He matrix, green squares correspond to Ar matrix, and blue diamond corresponds to p-$H_2$ measured at 2.8 K. Inset (a) denotes the sharp $\omega_H$ peaks for the small clusters. Size-reduction stiffens the H-O bonds with little disturbance to the dangling H-O bonds at 3700 cm$^{-1}$ in (b). (Reprinted with permission from [11, 53].)

Figure 10 feature the cluster-size-dependence of the calculated vibration spectra of $(H_2O)_N$ with respect to the ice-Ih phase. As expected, N reduction stiffens the $\omega_H$ from 3100 to 3650 cm$^{-1}$ and meanwhile softens the $\omega_L$ from 250 to 170 cm$^{-1}$ as the bulk water turns into dimers. The ∠O:H-O bending mode $\omega_{B1}$ (400–

1000 cm$^{-1}$) shifts to a slightly lower value, but the ∠H-O-H libration mode $\omega_{B2}$ (≈1600 cm$^{-1}$) remains unchanged [59].

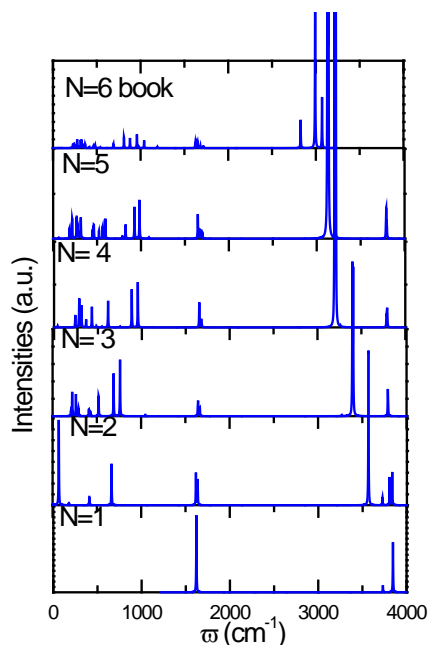

Figure 10. Cluster size dependence of the phonon $\omega$ relaxation in $(H_2O)_N$ clusters. Features corresponding to O:H-O bond segmental stretching and angle bending, as illustrated in **Error! Reference source not found.**. (Reprinted with permission from [46].)

N-reduction-stiffened $\omega_H$ in Figure 2b is consistent with spectroscopic measurements (Figure 9a). For instance, reduction of the $(H_2O)_N$ cluster from N = 6 to 1 stiffens the $\omega_H$ from 3200 to 3650 cm$^{-1}$ [10]. The skin $\omega_H$ of 3450 cm$^{-1}$ corrsponds to an effective cluster size of N = 2–3. Indeed, molecular undercoordination shortens and stiffens the H-O bond, and lengthens and softens the O:H nonbond consistently, which confirms the proposal of O:H-O bond cooperativity and the BOLS notion.

4.6     Potential paths of the O:H-O bond relaxing with cluster size

Lagrangian solution to the O:H-O bond oscillating dynamics [20] transforms the known segmental length and phonon frequency ($d_x$, $\omega_x$) for the H-O bond (x = H) and the O:H nonbond (x = L)[7] into their force constants and bond energies ($k_x$, $E_x$) turned out the potential paths for the O:H-O bond relaxing with $(H_2O)_N$ cluster size. Figure 11a shows that the presence of the Coulomb repulsion between electron pairs on adjacent oxygen atoms shifts both oxygen atoms along the O:H-O bond outward slightly from the initial posions denoted with blu dots with respect to the H atom coordination origin. Molecular undercoordination shifts both oxygen ions toward left of the O:H-O bond by different amounts. The H-O reduces from 0.99 to 0.97 Å and the O:H shifts from 1.66 to 1.92 Å when the N drops from 6 to 2.

The combination of Coulomb repulsion and molecular undercoordination not only reduces the molecular size ($d_H$) with enhanced intra-molecular interaction but also enlarges the molecular separation ($d_L$) with attenuated inter-molecular nonbond strength. The relaxation increases the H-O cohesive energy, as measured, from the bulk value of 3.97 eV [8], to the skin of 4.66 eV [6], and to the gaseous monomers 5.098 of $H_2O$ [60], 5.10 eV of $H_2O$ [61] and $D_2O$ [62] as well. The $E_L$ for N = 6 is around the bulk value of 0.095 eV [24]. The total energy gain Figure 11b with CN reduction in a linear dependence, which plays a role in the regelation. O:H-O bond tends form gain when the skin is subject to contact with recovery of molecular CN.

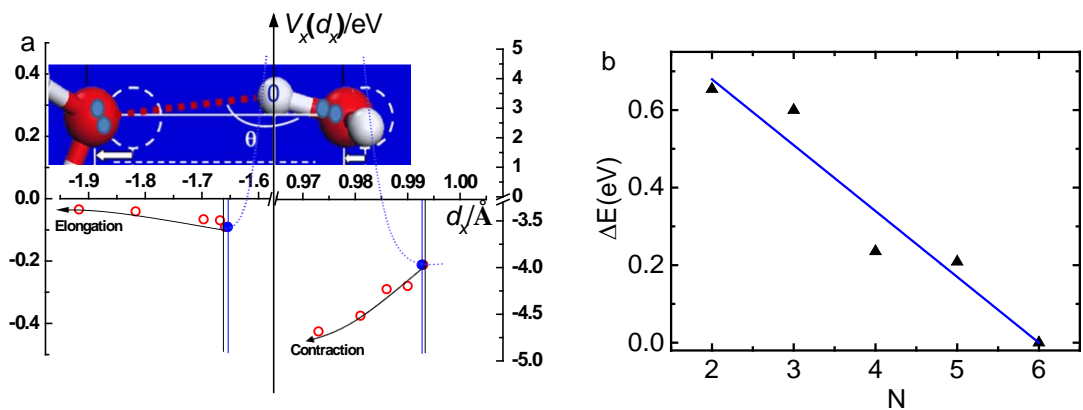

Figure 11. (a) Potential paths (red circles) for the O:H-O bond relaxing with cluster size N (r. to l.: N = 6, 5, 4, 3, 2) in the $(H_2O)_N$ clusters. O ions shift from the initial equilibrium (blue dots) to new equilibrium (red leftmost) with respect to H proton origin upon inter-oxygen Coulomb repulsion being involved. Molecular undercoordination reduces the $H_2O$ size ($d_H$) but increases their separations ($d_L$) with H-O bond stiffening and O:H nonbond softening. (b) The total energy shift, $\Delta(E_L + E_H)$, with N reduction.

5    Insight extension: Nanodroplet thermodynamics

5.1    Supercooling or superheating?

Undercoordinated water molecules are even more fascinating [4, 19, 34, 39, 42, 55, 63-66] than those fully coordinated in the bulk. Water nanodroplets undergo "supercooling" at freezing and "supercooling" at melting. Water droplets encapsulated in hydrophobic capillaries [67, 68] and ultrathin water films deposited on graphite, silica, and certain metals [65, 69-76] behave like ice at room temperature. The transition temperature for liquid formation ($T_m$) shifts from the bulk value of 273 K [21] up to 310 K for the skin [8]. A monolayer of ice melts at 325 K according to MD calculations [77] compared to around 310 K of the skin of bulk water [78]. Molecules at the air/water interface and those at the hydrophobic contacting interface performed the same as both are subjected to undercoordination.

The droplet size effect on $T_m$ and $T_N$ is often reffered to "superheating" at melting and "supercooling" at freezing. Supercooling, also known as undercooling [79], is the process of lowering the temperature of a liquid or a gas below its freezing point without it becoming a solid. Superheating is the opposite. Supercooled water occurs in the form of small droplets in clouds and plays a key role in the processing of solar and terrestrial radiative energy fluxes. Supercooled water is also important for life at subfreezing conditions for the commercial preservation of proteins and cells, and for the prevention of hydrate formation in nature gas pipelines.

??

Figure 12. Superheating and supercooling.

Melting point elevation is more apparent for molecules at the curved skin. Sum frequency generation (SFG) spectroscopy reveals that outermost two molecular layers are highly ordered at the hydrophobic contacts compared with those at a flat water–air interface [80]. Figure 13 shows the expected cluster size dependence of the $T_m$ and the O 1$s$ core level shift ($\Delta E_{1s}$) as a function of cluster size. As both $T_m$ and $\Delta E_{1s}$ are proportional to the covalent bond energy in the form of:

$T_{mN}/T_{mB} = \Delta E_{1sN}/\Delta E_{1sB} = E_{HN}/E_{HB} = (d_{HN}/d_{HB})^{-4}$. Subscript $B$ denotes the bulk.

One can derive from the plots that when the $N$ is reduced from a value of infinitely large to two, the $T_m$ will increase by 12% from 273 K to 305K, which explains why the ultrathin water films [69, 71-75, 81] or water droplets encapsulated in hydrophobic nanopores [67, 82] behave like ice at room temperature. The expected O 1$s$ energy shift ($C_z^{-4}$-1) of water clusters also agrees with the trend of the measurements. For instance, the O1$s$ core level shifts from 538.2 to 538.6 eV and to 539.8 eV, when the water cluster size is reduced from $N$ = 200 to 40 and to free water molecules [83, 84].

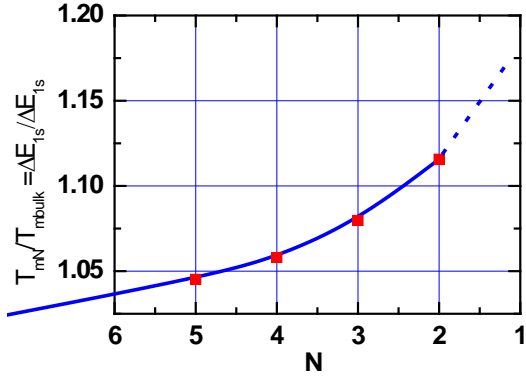

Figure 13 $N$-dependence of (a) the O—O distance, (b) the melting point, $T_{mN}$, (to $N = 2$ for dimers) and the O1$s$ core-level shift (to $N = 1$ for gas monomers) of (H$_2$O)$_N$ clusters based on DFT derived $d_{HN}/d_{HB}$ values and the expression of $T_{mN}/T_{mB} = \Delta E_{1sN}/\Delta E_{1sB} = E_{HN}/E_{HB} = (d_{HN}/d_{HB})^{-m}$, (m = 4 for covalent bond).

Amazingly, molecular undercoordination elavated $T_m$ is coupled with the depression of the homogeneous ice nucleation temperature ($T_N$). Figure 14a shows that the $T_N$ drops from the bulk value of 258 K [21] to 242 for 4.4 nm sized droplet, 220 K for 3.4 nm [85], 205 K for 1.4 nm [86] and 172 K for 1.2 nm sized droplets [87]. Freezing transition for clusters containing 1-18 molecules cannot be observed at temperature even down to120 K [88].

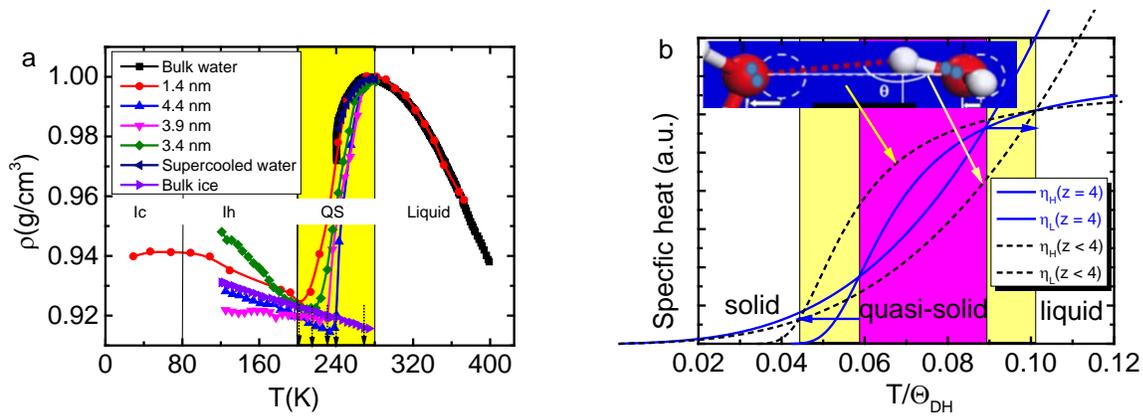

Figure 14. (a) Droplet size dependence of the homogeneous ice nucleation temperature ($T_N$) drops from the bulk value of 258 K [21] to 242 for 4.4 nm, 220 K for 3.4 nm [85], 205 K for 1.4 nm[86] and 172 K for 1.2 nm sized droplets [87]. Freezing transition for clusters containing 1-18 molecules cannot be observed at temperature down to120 K [88].The $\omega_x$ offsets the specific heat $\eta_x(T/\Theta_{Dx})$ through the Debye temperature $\Theta_{DH}(\omega_H)$. Superposition of the specific-heat $\eta_x(T/\Theta_{Dx})$ curves yields two intersecting temperatures $T_m$ and $T_N$ that form the boundaries of the quasi-solid phase. Molecular undercoordination (z

< 4) stretches $\eta_H(T/\Theta_{DH})$ by raising the $\Theta_{DH}(\omega_H)$ and depresses the $\eta_L(T/\Theta_{DL})$ by lowering the $\Theta_{DL}(\omega_L)$, which disperses the intersecting temperatures in opposite directions. Therefore, nanodroplets, nanobubbles, and water ice skins undergo simultaneously $T_N$ depression and $T_m$ elevation and the extent of dispersion varies with the fraction of undercoordinated molecules of the object.

Supercooling/heating is often confused with freezing/melting point depression/elevation. Freezing point depression occurs when a solution can be cooled below the freezing point of the corresponding pure liquid due to the presence of the solute or droplet size reductiomn; an example of this is the freezing point depression that occurs when sodium chloride is added to pure water. In fact, $T_m$ elevation is different from "superheating" and $T_N$ depression is different from "supercooling". The former is intrinsic and the latter is process dependent.

Why does droplet size efect on the $T_m$ and the $T_N$?

5.2    Size depressed melting point in general

Generally, melting a specific atom inside a normal substance requires heat that is a fraction of its cohesive energy, $E_C = zE_z$, i.e., the sum of bond energy $E_z$ over its coordination neighbors (z or CN). The $T_m$ of a solid changes with the solid size because of the skin atomic undercoordination and the varied fraction of undercoordinated atoms in the skin [16]. However, for water and ice, the presence of the critical temperatures at 273 K ($T_m$) and 258 K ($T_N$), see Figure 14, for transiting the bulk liquid into the quasisolid and then into solid [21] indicates that a quasisolid (or quasiliquid) phase exists in this temperature regime. Traditionally, the quasisolid phase is absent from existing phase diagrams.

5.3    Quasisolid phase in water

Firstly, one has to consider the specific-heat per bond $\eta(T/\Theta_D)$ in Debye approximation when dealing with the thermodynamic behaviour of a substance from the atomistic point of view. The specific-heat is regarded as a macroscopic quantity integrated over all bonds of the specimen, which is also the amount of energy required to raise the temperature of the substance by 1 °C. The specific-heat per bond is obtained by dividing the bulk specific-heat by the total number of bonds involved [89]. For other usual materials, one bond represents all on average, and therefore, the thermal response of all the bonds are the same, without any discrimination in responding to thermal excitation [90].

However, for water ice, the representative hydrogen bond (O:H-O) is composed of two segments that have their respective specific heat. The strong disparity between the specific-heat $\eta_x(T/\Theta_{Dx})$, as illustrated in Figure 14b, makes water perform differently from other normal substance with a unique $\eta(T/\Theta_D)$. Parameters characterize the $\eta_x(T/\Theta_{Dx})$ include the Debye temperature $\Theta_{Dx}$ and the thermal integration of the $\eta_x(T/\Theta_{Dx})$. The $\Theta_{Dx}$, which is lower than the $T_{mx}$, determines the rate of the $\eta_x(T/\Theta_{Dx})$ curve reaching saturation. The $\eta_x(T/\Theta_{Dx})$ curve of the segment with a relatively lower $\Theta_{Dx}$ value will rise to saturation quicker than the other segment. The $\Theta_{Dx}$ is proportional to the characteristic frequency of vibration $\omega_x$, acccording to Einstein's relastion: $\hbar\omega_x = k\Theta_{Dx}$, with $\hbar$ and $k$ being constant.

On the other hand, the integral of the $\eta_x(T/\Theta_{Dx})$ curve from 0 K to the $T_{mx}$ is proportional to the cohesive energy $E_x$ per segment [89]. The $T_{mx}$ is the temperature at which the vibration amplitude of an atom/molecule expands abruptly to more than 3% of its diameter irrespective of the environment or the size of a molecular cluster [91, 92].

Thus, with the known values of $\omega_L \sim 200$ cm$^{-1}$ for O:H stretching and $\omega_H \sim 3200$ cm$^{-1}$ for H-O stretching [21], $\Theta_{DL} = 198$ K $< 273$ K ($T_m$), $E_L = 0.095$ eV [24], and $E_H = 3.97$ eV [7], one can estimate $\Theta_{DH} \approx 16 \times \Theta_{DL} \approx 3200$ K and $T_{mH} >> \Theta_{DH}$ from the following,

$$\begin{cases} \Theta_{DL}/\Theta_{DH} \approx 198/\Theta_{DH} \approx \omega_L/\omega_H \approx 200/3200 \sim 1/16 \\ \left(\int_0^{T_{mH}} \eta_H dt\right) / \left(\int_0^{T_{mL}} \eta_L dt\right) \approx E_H/E_L \approx 4.0/0.1 \sim 40 \end{cases}$$

(2)

The $\eta_L$ ends at $T_{mL} = 273$ K and the $\eta_H$ ends at $T_{mH} \sim 3200$ K, which means that the area covered by the $\eta_H$ curve is 40 times that covered by the $\eta_L$ curve.

Secondly, a superposition of these two $\eta_x$ curves yields two intersecting temperatures that divide the full temperature range into water phases of liquid, quasisolid, and solid with different $\eta_L/\eta_H$ ratios, see Figure 14b. In the liquid and in the solid phase ($\eta_L/\eta_H < 1$), the O:H nonbond contracts more than the H-O expands at cooling, resulting in the cooling densification of water and ice [85, 86]; in the quasisolid phase, the O:H and the H-O swap roles ($\eta_H/\eta_L < 1$), the H-O contracts less than the O:H expands at cooling, so the $\Delta d_{OO} > 0$ and water in quasisolid phase become less dense as it cools, which is responsible for ice floating [21]. At the quasisolid phase boundaries ($\eta_H/\eta_L = 1$), $\Delta d_L$ and $\Delta d_H$ change sign, which correspond to density extremes. Ideally, the $T_m$ corresponds to the maximal density at 4 °C liquid and the $T_N$ the minimal density [85, 86, 93].

Molecular undercoordination does sttifens the H-O phonon $\omega_H$ and meanwhile softens the O:H phonon $\omega_L$, intrinsically [46], and shifts the O 1s binding energy positively as well. The $\omega_H$ shifts from 3200 to 3650 cm$^{-1}$ when the N of the $(H_2O)_N$ cluster drops from 6 to 1 [10-12] and and meanwhile softens the $\omega_L$ from 260 to 170 cm$^{-1}$ as the bulk water turns into dimer [10]. Furthermore, molecular undercoordination shifts the O 1s energy level more deeply from the bulk value of 536.6 eV to 538.1 eV and 539.7 eV when bulk water is transformed into skin or into gaseous molecules [23, 43, 44].

5.4    Phase boundary dispersion

One can imagine what will happen to the phase boundaries by raising the $\Theta_{DH}$ and meanwhile lowering the $\Theta_{DL}$. The $\eta_L$ will saturate quicker and the $\eta_H$ slower than they were in the bulk standard. This process will raise the $T_m$ and lower the $T_N$, as illustrated in Figure 14b. According to the relationship in eq (2) ($\Theta_{DL}$= 198 K, $\Theta_{DH}$ = 3200 K) and data in Figure 15, the $\Theta_{Dx} \approx \omega_x$ in absolute values though the calculated $\omega_D$ is subject to modification with respect to measurements and $\omega_L$ is experimentally hardly avaiable[21]. One can estimate $\Theta_{DL}$ =198 (for bulk)/260(calculated bulk)×195(calculated cluster) = 149 K and $\Theta_{DH}$ =3200/3200×3550 = 3550 K, for N = 2.

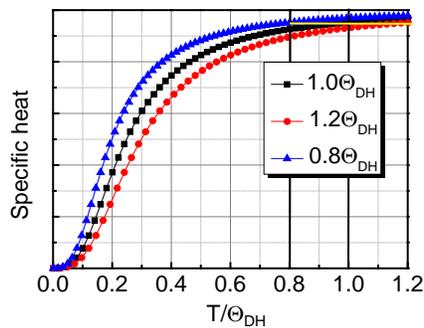

Figure 15. $\Theta_{DH}(\omega_H)$ relaxation disperses the $\eta_H$ curve.

With the known bulk values of $\Theta_{DL}$ = 198 K, $T_m$ = 273 K., $T_N$ = 258 K, and the respective $\omega_x$ and $E_x$ (in Table 2) one can estimate the cluster size dependence of the $\Theta_{Dx}$, $T_m$, and $T_N$ using the following relationships [16]:

$$\begin{cases} \Theta_{Dx} \propto \omega_x \\ T_{N,m} \propto E_{L,H} \end{cases}$$

Numerical reproduction of the $T_m(P)$ profiles indicates that the $T_m$ is proportional to $E_H$ and Figure 14a suggests that the $T_N$ be proportional to $E_L$. In order to minimize calculation artifacts, a modification of the $\omega_L(N)$ curve in Figure 2b is made with respect to the measured value of 220 cm$^{-1}$ for bulk water and to that the calculated $\omega_x$ matches the measured value at N = 2. This modification improves the precision of estimating cluster size dependence of the $\Theta_{DL}$. As featured in Table 2, the estimated N-dependent $T_m$ and $T_N$ agree with trends of observations.

Table 2. DFT-derived segmental length $d_x$, ∠O:H-O containing angle θ, and phonon frequency $\omega_x$ for $(H_2O)_N$ clusters[7]. Presented are also N dependence of the Debye temperatures $\Theta_{Dx}$, freezing temperature $T_N$, and melting point $T_m$ estimated herewith, # indicates the corrected $\omega_L$ with respect to that 220 cm$^{-1}$ for bulk water and to that the calculated $\omega_H$ matches the measured value at N = 2.

|  | Monomer | Dimer | Trimer | Tetramer | Pentamer | hexamer | bulk[21] |
|---|---|---|---|---|---|---|---|
| N | 1 | 2 | 3 | 4 | 5 | 6 | Ih |
| $d_H$(Å) | 0.969 | 0.973 | 0.981 | 0.986 | 0.987 | 0.993 | 1.010 |
| $d_L$(Å) | - | 1.917 | 1.817 | 1.697 | 1.668 | 1.659 | 1.742 |
| θ (°) | - | 163.6 | 153.4 | 169.3 | 177.3 | 168.6 | 170.0 |
| $\omega_L$ (cm$^{-1}$) | - | 184 | 198 | 229 | 251 | 260 | - |
| $\omega_L$ (cm$^{-1}$)# | - | 184 | 190 | 200 | 210 | 218 | 220 |
| $\omega_H$(cm$^{-1}$)$^{50-52,55}$ | 3650 | 3575 | 3525 | 3380 | 3350 | 3225 | 3150 |
| $\Theta_{DL}$(K) | - | 167 | 171 | 180 | 189 | 196 | 198[24] |
| $\Theta_{DH}$(K) | 3650 | 3575 | 3525 | 3380 | 3350 | 3225 | 3150 |
| $E_L$(meV) | - | 34.60 | 40.54 | 66.13 | 69.39 | 90.70 | 95 |
| $T_N$(K) | - | 94 | 110 | 180 | 188 | 246 | 258 |
| $E_H$(eV) | 5.10 | 4.68 | 4.62 | 4.23 | 4.20 | 3.97 | 3.97 |
| $T_m$(K) | - | 322 | 318 | 291 | 289 | 273 | 273 |

*Experimentally observed $T_m$ elevation and $T_N$ depression:
$T_m$ = 325 K (monolayer)[77]; 310 K (skin of bulk)[8].

$T_N$ = 242 K (4.4 nm droplet)[85]; 220 K (3.4 nm droplet)[85]; 205 K (1.4 nm droplet)[86]; 172 K (1.2 nm droplet)[87]; <120 K (1-18 molecules)[88].

These phase boundary dispersivity is responsible for the thermodynamic behavior of water droplets and gas bubbles, particularly at the nanometer scales. These systems of undercoordinated molecular dominance have far-reaching physical, chemical, and biological effects [94] because molecular undercoordination induced unusual bond-electron-phonon behavior, as afore discussed. They are hardly destroyed and thermally much more stable than bubbles at the millimeter scale [95]. Water nanodroplets and nanobubbles do follow the trend of $T_m$ elevation and $T_N$ depression because of the dominant fraction of undercoordinated skin molecules. Droplet size reduction raises the $\Theta_{DH}(\omega_H)$ and stretches the $\eta_H(T)$ curve and meanwhile, lowers the $\Theta_{DL}(\omega_L)$ and compresses the $\eta_L(T)$ curve, which disperses the extreme-density temperatures. A bubble is just the inversion of a droplet; a hollow sphere like a soap bubble contains two skins – the inner and the outer. Both skins are in the supersolid phase and the volume fraction of such supersolid phase over the entire liquid-shell volume is much greater than simply a droplet. Therefore, bubbles demonstrate more significantly the supersolidity nature – elastic, hydrophobic, and thermally stable, which makes bubbles mechanically stronger and thermally more stable [8].

## 6    Summary

Thus, a hybridization of the H-O bond contraction [15, 96-98] dual polarization [99, 100] of the segmented hydrogen bond notion [10, 101] clarifies the origin of the observed length scale, binding energy, phonon frequency, thermal stability and th ethermodynamics of water molecules with fewer-than-four neighbors such as molecular clusters, hydration shells, snowflakes, and surface skins of liquid water. This notion also reconciled the anomalies of O—O expansion, O1s electron densification and entrapment, surface electron polarization, high-frequency phonon stiffening, and the ice like and hydrophobic nature of such undercoordinated water molecules. Agreement between numerical calculations and experimental observations has verified our hypothesis and predictions:

1)  Under-coordination-induced contraction of the H-O bond and inter-electron-pair repulsion driven O:H elongation dictate the unusual behaviour of water molecules in the nanoscale O:H-O networks and in the skin of water.

2) The shortening of the H-O bond raises the density of the core and bonding electrons in the under-coordinated molecules, which in turn polarizes the nonbonding electron lone pairs on oxygen.
3) The stiffening of the H-O bond increases the O1s core-level shift, causes the blue-shift of the H-O phonon frequency, and elevates the melting point of water molecular clusters, surface skins, and ultrathin films of water.
4) Under-coordinated water molecules could form an ice-like, low-density phase that is hydrophobic, stiffer, and thermally more stable than the bulk water[63, 64]
5) Undercoordination-induced O:H-O relaxation results in the supersolid phase that is elastic, hydrophobic, thermally more stable, and less dense, which dictates the unusual behaviour of water molecules at the boundary of the O:H-O networks or in the nanoscale droplet.
6) H-O bond contraction densifies and entraps the core and bonding electrons; H-O bond stiffening shifts positively the O1s energy, the $\omega_H$ and the $T_m$ of molecular clusters, surface skins, and ultrathin films of water.
7) The dual polarization makes the skins hydrophobic, viscoelastic, and frictionless.
8) H-O bond contraction elevates the melting point and O:H nonbond elongation depresses the freezing temperature of water droplets and bubbles of which the undercoordinated molecules become dominant.

**3**: 1455-1460.